\documentclass[aps,prb,twocolumn,unsortedaddress,showpacs,showkeys]{revtex4-1}

\usepackage{natbib}
\usepackage{graphicx}
\usepackage{dcolumn}
\usepackage{multirow}

\newcommand{\mlr}[1]{\multirow{2}{*}{#1}}
\newcommand\T{\rule{0pt}{2.5ex}}
\newcommand\B{\rule[-1.2ex]{0pt}{0pt}}

\begin{document}


\title{First-principles atomistic Wulff constructions for gold nanoparticles}


\author{Georgios D. Barmparis}
\affiliation{Department of Materials Science and Technology, University of Crete, 71003 Heraklion, Crete, Greece}
\affiliation{Institute of Electronic Structure \& Laser, Foundation for Research and Technology-Hellas, Heraklion, Crete, 71110, Greece}
\email[Electronic Address: ]{barmparis@materials.uoc.gr}

\author{Ioannis N. Remediakis}
\affiliation{Department of Materials Science and Technology, University of Crete, 71003 Heraklion, Crete, Greece}
\email[Electronic Address: ]{remed@materials.uoc.gr}
\homepage[Group Web Page: ]{http://theory.materials.uoc.gr}
\date{\today}

\begin{abstract}
We present a computational study for the equilibrium shape of gold nanoparticles. By linking extensive
quantum-mechanical calculations, based on Density-Functional Theory (DFT) to Wulff construction, we predict 
equilibrium shapes that are in good agreement with experimental observations. We discuss the effect of the interactions between a 
nanoparticle and the encapsulating material on the equilibrium shape. As an example, we calculate adsorption of CO on several 
different Au($hkl$) and use the results to explain the experimentally observed shape change of Au nanoparticles.
\end{abstract}


\maketitle

Bulk gold is the noblest of all metals\cite{Hammer:Why-Gold:1995}, as demonstrated by delicate
gold jewels manufactured several millennia ago which are found
intact in excavations. On the other hand, catalysts that include
oxide-supported gold nanoparticles were found to efficiently
oxidize CO at room temperature \cite{Haruta:Gold:1989, Valden:Onset:1998, Uchiyama:Systematic:2011}; Au is by far the best such catalyst \cite{Falsig:Trends:2008}.
Among the key factors that determine the efficiency of Au catalysts is the shape of Au nanoparticles, in particular the 5- and 6-fold coordinated atoms at its corners \cite{Lopez:Catalytic:2002,Remediakis:CO-Oxidation:2005,Remediakis:CO-oxidationGold:2005}.

The shape of Au nanoparticles has a key role in every aspect of
their functionality, from sensing \cite{ ISI:000173708300001}
and biolabeling applications \cite{Boisselier:Gold:2009} to
plasmonics \cite{Lal:Nano-optics:2007} and photonics
\cite{Lu:Synthesis:2002}. In optoelectronics, quantum leaps
between electronic states transform light into electricity and vice versa. 
The probability
of such a transition depends on the density of states and the
dipole matrix elements according to Fermi's golden rule. For a
given size, both wavefunctions and energies depend critically on
the nanoparticle shape. For example, the lowest excitation
energy for a cubic nanoparticle is 10\% higher than that of a
spherical nanoparticle of the same volume \footnote{For cube of
edge $a$, $\Delta E=\frac{\hbar^2\pi^2}{2ma^2}$; for a sphere of
radius $r$, $\Delta
E=\frac{\hbar^2(\chi_{11}-\chi_{10})}{2mr^2}$, where
$\chi_{11},\chi_{10}$ are the first two roots of the spherical
Bessel function $j_1(x)$}. 

Gold nanoparticles are often found in their
equilibrium shape. 
This is a polyhedron enclosed by faces of
various $(hkl)$ crystal orientations such that the total surface
energy,
\begin{equation} 
\sum_{hkl} A_{hkl} \gamma_{hkl}, 
\label{Equ:energy}
\end{equation}
is minimum. $A_{hkl}$ is the total area of faces parallel to $(hkl)$ plane of the crystal and $\gamma_{hkl}$ is the surface tension, i.e. the energy required to create a surface of unit area that is parallel to the ($hkl$)
plane of the crystal. In order to predict equilibrium shape, one needs
calculations of surface tensions for many different $(hkl)$.
Several such calculations exist in the literature either based on empirical
potentials \cite{Wen:Surface:2007} or limited to Miller indexes of 0 and 1
\cite{Vitos:The-surface:1998}, or using quantum-mechanics for low-index faces
and empirical models for higher indexes \cite{Galanakis:Broken-bond:2002}. An accurate  and systematic calculation of all high-index Au surfaces is missing.

The equilibrium shape is often found to change upon exposure to some interacting environment.  As Au nanoparticles are used in CO oxidation catalysis, CO gas is the ideal candidate to test this idea. Changes to shapes of higher sphericity upon exposure to CO gas have been observed both experimentally \cite{Ueda:First:2008} and theoretically\cite{mckenna09}. The interface tension of a metal in equilibrium with a gas is found to depend on the surface tension, the adsorption energy and the coverage of adsorbates Eq. (\ref{equ:gammaInt}). In order to predict the equilibrium shape in an interacting environment using the Wulff construction method, it is necessary to have a systematic calculation of adsorption energies for all relevant $(hkl)$ surfaces.  

The Wulff construction has been used to predict equilibrium shapes in a
variety of systems. Wulff polyhedra are often employed in observations and
models for nanomaterials including Cu catalysts
\cite{Clausen:Wetting:1994,Hansen:Atom-resolved:2002}, or semiconductors
\cite{Muller:Equilibrium:2000}. In the past decade, Wulff shapes employing
surface tensions from first-principles calculations were used for the
successful prediction of the shape of nanoparticles, including interactions
with their environment
\cite{barnard04,barnard05,Hadjisavvas:Insights:2006,Kopidakis:Atomic:2007,mittendorfer07,Soon:Morphology:2008,Shi:Shape:2008}.
In the context of ammonia-synthesis catalysis, an {\em ab initio}
determination of a nanoparticle shape was used as a first step in the
creation of a virtual nano-catalyst \cite{Honkala:Ammonia:2005,hellman06,hellman09}.
In that work, the Wulff polyhedron was filled with atoms in order to create
a realistic nanoparticle. The advantage of this method was that it allowed
for detailed analysis of the atomic positions, making it possible to
calculate structural quantities such as the number of active sites. This
virtual nano-catalyst was used in other similar reactions, such as ammonia
decomposition\cite{hellman09}. Here, we expand this methodology by including all possible $(hkl)$ orientations. Moreover, we take into account changes in shape that may be induced by interactions between the nanoparticle and its environment.  We apply our method to supported gold nanoparticles, a system of high technological importance. 

The paper is organised as follows: in Section I, we briefly review  Wulff's theory regarding the equilibrium shape. In Section II, we present a systematic calculation of the surface tension for every Au($hkl$) with Miller indexes up to 4. In  section III, we use these surface tensions to create atomistic models for Au nanoparticles of sizes up to 70 nm, and analyse their structural properties, such as the concentration of active sites.  In Section IV, we generalize our methodology for nanoparticles in interacting environment. We provide a simple formula that relates the interface tension to the surface tension and adsorption energy. We calculate the minimum adsorption energy of CO on every Au surface with Miller indexes up to 3, and use these results to calculate the change in equilibrium shape of Au particles upon exposure to CO gas. We summarize our results in Section V.

\section{The Wulff construction}

The concept of "equilibrium shape" was postulated by Gibbs in the late 19th
century. Under thermodynamic equilibrium, a given quantity of matter will
attain a shape that minimizes the total surface energy of the system. 
More than a century ago, mineralogist G. Wulff
proposed  that the shape that minimizes Eq. (\ref{Equ:energy}) is such that the distance of each face from the center is proportional
to the surface tension of the respective $(hkl)$ surface \cite{Herring:Some:1951}:
\begin{equation}
d_{hkl}\sim \gamma_{hkl}. 
\label{eq:wulfftheorem}
\end{equation} 
One begins the Wulff construction by drawing up a
plane (for example, (111)) at a distance $d_{111}$ from the origin followed by
planes parallel to ($hkl$) at distances $d_{hkl} =
d_{111}\gamma_{hkl}/\gamma_{111}$. The equilibrium shape will be
the polyhedron enclosed by these planes, having thus the following properties:
\renewcommand{\labelenumi}{(\alph{enumi})}
\begin{enumerate}
\item The shape depends on ratios between surface
tensions, and not their absolute values. 
\item $(hkl)$ planes with
high surface tension (usually high-indexed ones) will be drawn
at greater distances and are therefore less likely to appear in
the equilibrium shape.
\item  Being steeper, high-index
faces are usually hidden behind low-index ones, and tend to occupy smaller areas in the
equilibrium shape even if $\gamma_{hkl}$ is low. 
\item The extra
energy associated with the formation of edges between two
surfaces is not taken into account.
\item The Wulff polyhedron belongs 
to the same point group as the crystal structure of the material.
\end{enumerate}

In addition to Wulff construction, there exist other methods for the study of nanoparticles.  Advances in computers  allow for the direct simulation of nanoparticles of large sizes using empirical potentials, as done for example by McKenna \cite{mckenna09}. In that work, a large number of different shapes are tested to find the lowest-energy one. The Wulff construction is complimentary to that method. Here, we use Wulff construction coupled to first-principles calculations of surface tensions. This method offers a systematic, easy-to-follow recipe for the construction of atomistic models of nanoparticles. 


\section{Surface tension of gold surfaces}

\begin{figure}
\includegraphics[width=\columnwidth]{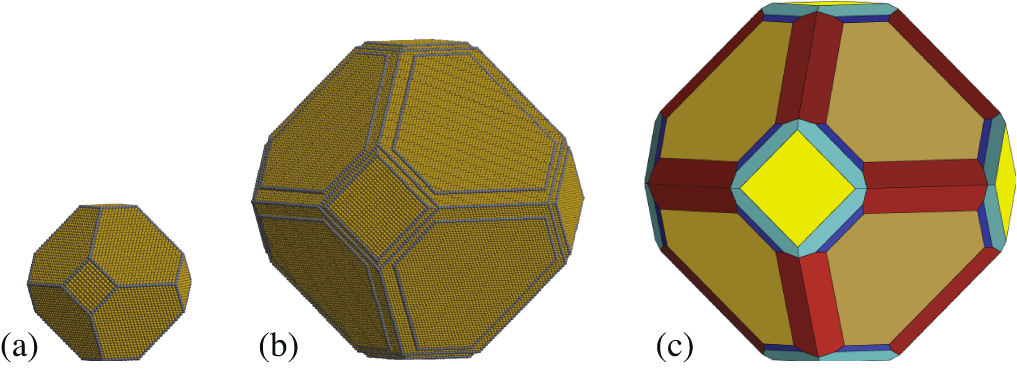}
\caption{(Color online) Typical calculated Au nanoparticles for various sizes, $d$: (a) $d=12.1$ nm (b) $d=27.2$ nm (c) $d\to\infty$. 
In (a) and (b), step and kink atoms are shown in darker color. In (c), different colors correspond to different kinds of surfaces. (c) 
was created using Wulffman \cite{Roosen:Wulffman::1998}.
\label{fig:goldNanoparticles}}
\end{figure}

As we are interested in relatively large Au nanoparticles, we limit our
study to nanoparticles where Au atoms far from the surfaces are in the
ideal fcc lattice. This is observed in simulations of large clusters
\cite{Cleveland:Structural:1997}, although small gold clusters may have
structures very different from fcc \cite{baletto05}. We begin by calculating
the surface tension, $\gamma_{hkl}$, of Au by simulations of periodic
($hkl$) slabs using Density-Functional Theory (DFT). We use the open-source
Dacapo/ASE suite (https://wiki.fysik.dtu.dk). We use a plane wave basis
with 340 eV cut-off. The core electrons are treated with Vanderbilt
non-local ultrasoft pseudopotentials\cite{vanderbilt90}. The Brillouin zone
of the (111)-(1$\times$1) surface is modelled by a (10$\times$10$\times$1)
Monkhorst-Pack grid of $\vec{k}$-points; number of $\vec{k}$-points in
other surfaces is calculated in proportionality to the (111) cell. We use
the generalized gradient approximation (GGA) Perdew-Wang
exchange-correlation functional PW91 for the clean surfaces and the revised
Perdew-Burke-Ernzerhof functional RPBE for the surfaces covered with CO
since this one gives better adsorption energies
\cite{Hammer:Improved:1999}. For each set of calculations, we use the
theoretical lattic constant which is found to be 4.22 \AA\ for RPBE and
4.18 \AA\ for PW91, very close to the experimental value of 4.08 \AA. There
is a general trend to slightly overestimate the lattice constant of noble
metals using GGA \cite{fuchs98}. We model all $(hkl)$ surfaces of fcc Au
with indexes up to 4. Even for CO-covered nanoparticles, no $(4kl)$
surfaces are observed in the Wulff construction; for this reason we do not
consider $(5kl)$ surfaces in this work. Atoms in the top two layers from
each side are allowed to relax, while subsequent slabs are separated by 12
\AA\ ~of vacuum. Slab thickness is chosen independently for each ($hkl$) slab
until the surface tension converges within 0.01 J/m$^2$. The surface
tension, $\gamma_{hkl}$ is derived from 
\begin{equation} E_{slab} = N
E_{bulk}+2A\gamma_{hkl},
\label{equ:Eslab} 
\end{equation} 
where $N$ is the number of atoms in the slab, $E_{slab}$ is the total energy of the slab,
$E_{bulk}$ is the energy per atom in bulk Au and $A$ is the area parallel
to $(hkl)$.

\begin{table}
\begin{center}
\begin{tabular}{cccccl}
\hline
\hline  
& This work & calc\cite{Wen:Surface:2007} &
calc\cite{Galanakis:Broken-bond:2002} & calc\cite{Vitos:The-surface:1998}  & $E_{ads}$ (eV) \\
\hline

$\gamma_{100}/\gamma_{111}$   & 1.23 & 1.11  & 1.15  & 1.27  & -0.25 (b)\\
$\gamma_{110}/\gamma_{111}$   & 1.29 & 1.24  & 1.22  & 1.33  & -0.36 (b)\\ 
$\gamma_{210}/\gamma_{111}$   & 1.33 & 1.31  & 1.29  &          & -0.49 (t) \\
$\gamma_{211}/\gamma_{111}$   & 1.17 & 1.19  & 1.18  &          & -0.34 (t)\\
$\gamma_{221}/\gamma_{111}$   & 1.14 & 1.16  & 1.15  &          & -0.35 (t)\\
$\gamma_{310}/\gamma_{111}$   & 1.31 & 1.28  & 1.28  &          & -0.51 (t)\\
$\gamma_{311}/\gamma_{111}$   & 1.26 & 1.24  & 1.22  &          & -0.32 (b)\\
$\gamma_{320}/\gamma_{111}$   & 1.36 & 1.30  & 1.28  &          & -0.48 (t)\\
$\gamma_{321}/\gamma_{111}$   & 1.25 & 1.26  & 1.23  &          & -0.50 (t)\\
$\gamma_{322}/\gamma_{111}$   & 1.11 & 1.13  & 1.12  &          & -0.34 (t)\\
$\gamma_{331}/\gamma_{111}$   & 1.18 & 1.21  & 1.19  &          & -0.34 (t)\\
$\gamma_{332}/\gamma_{111}$   & 1.07 & 1.11  & 1.11  &          & -0.35 (t) \\
$\gamma_{410}/\gamma_{111}$   & 1.32 & 1.25  & 1.26  &          & -0.49$^f$\\
$\gamma_{411}/\gamma_{111}$   & 1.27 & 1.23  & 1.22  &          & -0.34$^f$ \\
$\gamma_{421}/\gamma_{111}$   & 1.32 & 1.29  & 1.26  &          & -0.49$^f$ \\
$\gamma_{430}/\gamma_{111}$   & 1.34 & 1.29  & 1.27  &          & -0.49$^f$ \\
$\gamma_{431}/\gamma_{111}$   & 1.27 & 1.27  & 1.25  &          & -0.49$^f$ \\
$\gamma_{432}/\gamma_{111}$   & 1.19 & 1.20  & 1.18  &          & -0.49$^f$ \\
$\gamma_{433}/\gamma_{111}$   & 1.09 & 1.09  & 1.09  &          & -0.34$^f$ \\
$\gamma_{441}/\gamma_{111}$   & 1.22 & 1.22  & 1.21  &          & -0.34$^f$ \\
$\gamma_{443}/\gamma_{111}$   & 1.06 & 1.09  & 1.08  &          & -0.34$^f$ \\
\hline
\end{tabular}
\caption{Ratios of surface tensions of Au in comparison to other calculations. The last column contains the adsorption energy of CO at low coverage on the same $(hkl)$ surface, and the adsorption geometry. b=bridge adsorption, t=on-top adsorption. $^f$: energies calculated from the linear fit shown in Fig. \ref{fig:CO_Eads}.}
\label{tab:surfaceTensionTable}
\end{center}
\end{table}


The results are summarized in Table \ref{tab:surfaceTensionTable}.
Interestingly, the ratio of surface energies of different cells is very
close to the ratio of the areal density of cleaved bonds
\cite{Galanakis:Broken-bond:2002}. 
 This is another example of the unique chemistry of Au \cite{Hammer:Why-Gold:1995}: Au atoms
have a closed $d$-shell and have the least preference for directional bonds
in the entire periodic table. The calculated absolute value for
$\gamma_{111}$ is 0.69 J/m$^2$, very close to 0.64 J/m$^2$ reported by Wen
and Zhang \cite{Wen:Surface:2007} and within the same order of magnitude as
the values reported by state-of-the-art relativistic all-electron
calculations \cite{Galanakis:Broken-bond:2002,Vitos:The-surface:1998}. The
nanoparticle shape depends only on ratios between surface energies. As
shown in Table \ref{tab:surfaceTensionTable}, our results for the ratios
between surface tensions agree with more detailed calculations within 5\% or
less.
   
\begin{table*}
\begin{center}
\begin{tabular}{ccccccccc}
\cline{1-9}
\T \B Shape & d (nm) & N$_{corner}$ & N$_{edge}$ & N$_{surf}$ & N$_{tot}$ & Area (nm$^2$) & Volume (nm$^3$) & Faces \\
\cline{1-9}
\T \B \mlr{(a)} & \mlr{12.12} & \mlr{24} & \mlr{444} & \mlr{5208} & \mlr{42925} & \mlr{438} & \mlr{730} & (111), (100) \\
& & & & & & & & (86\%), (14\%)\\
\cline{1-9}
\T \B \mlr{(b)} & \mlr{27.17} & \mlr{96} & \mlr{2832} & \mlr{25998} & \mlr{473550} & \mlr{2275} & \mlr{8439} & (111), (332), (211), (100) \\
& & & & & & & & (58\%), (16\%), (15\%), (11\%) \\
\cline{1-9}
\T \B (c)  & 0.31N$^{0.34}$ & 144 & 6N$^{0.40}$ & 3N$^{0.68}$ & N & 0.3N$^{0.67}$ & 0.02N & (111),(332),(211),(100),(322) \\
\cline{1-9}
\end{tabular}
\end{center}
\caption{Characteristic data for typical nanoparticles shown in Fig. 1. Shape (a) is typical for particles up to $d=16.3$ nm in diameter, shape  (b) is typical for larger particles and (c) presents fitted values from over a hundred different particles with $d<$ 16.3 \AA. $N_{corner}$, $N_{edge}$ and $N_{surf}$ are the total number of atoms at vertices, edges and faces of the nanoparticle, respectively; $N_{total}$ is the total number of atoms. $A$ is the total surface area and $V$ the volume of the nanoparticle. $(hkl)$ are the appearing surfaces in the shape with the percentage of the total area they occupy.}
\label{tab:nanoData1}
\end{table*}

\section{Au nanoparticles in non-interacting environment}

The Wulff construction for Au is shown in Fig. \ref{fig:goldNanoparticles}(c). 
It contains 144 vertices and 86 faces of 5 different kinds: (111), (100), (332), (211) and
(322) in order of total area.

To construct atomistic models for nanoparticles, we start from a large fcc
crystal. As (111) has the lowest surface tension, we begin by choosing the
number of (111) layers. This determines the distance of (111) plane from
the center of the nanoparticle, $d_{111}$, and, consequently, the
nanoparticle size. For other faces, we use Eq. (\ref{eq:wulfftheorem}), with
the calculated values of $\gamma_{hkl}$ and cut the crystal at the correct
distances $d_{hkl}$. We calculate the equation of the plane defined by
every set of three surface atoms, and make sure that only faces consistent
with the Wulff construction appear on the nanoparticle. We consider about 30000 different
nanoparticles with diameters ranging from 1.7 nm to more than 100 nm. 

At small sizes, some faces might not be large enough to accommodate a single
atom, let alone a unit cell of this. Very small
nanoparticles expose only (111) and (100) faces; in
particular, our simulated 459-atom nanoparticle is identical to
the one found from simulations and X-ray experiments
\cite{Cleveland:Structural:1997}. In all cases, the shape resembles a truncated octahedron consisting mainly of (111) and 
(100) faces, with their edges decorated by several $(hkl)$ faces with indexes up to 3.
For diameters up to 16.3 nm, we find only (111) and (100) faces; as the nanoparticle grows in size, different ($hkl$) orientations start to appear. The thermodynamic limit, shown in
\ref{fig:goldNanoparticles}(c) is reached at diameters of the order of 100 nm.

Geometrical features or typical nanoparticles are shown in Table
\ref{tab:nanoData1}. The area of nanoparticles is calculated analytically
using the coordinates of vertices; their volume is obtained by numerical
integration. By fitting over a hundred particles of different diameters, we
provide scaling relations of various properties with total number of atoms,
in accordance with atom-counting models for nanoparticles
\cite{Lopez:On-the-origin:2004, Galanis:Mechanical:2010}.
%
%
%
%

\section{Au nanoparticles in  interacting environment}

The equilibrium shape of nanoparticles that interact with their
environment can be found by means of a Wulff construction
based on interfacial tensions,
$\gamma^{int}_{hkl}$, between Au and its environment instead of
surface tensions, $\gamma_{hkl}$. It turns out that the two
are related by a simple formula:
\begin{equation}
\gamma^{int}_{hkl} = \gamma_{hkl} + \theta \frac{E_{ads}}{A_{at}}, 
\label{equ:gammaInt}
\end{equation} 
where $\theta$ is the coverage (number of metal-adsorbate bonds over number
of surface atoms), $A_{at}$ is total surface area per metal atom and
$E_{ads}$ the adsorption energy, defined as the excess energy
per molecule of the system compared to isolated Au
surface and isolated encapsulating material. Eq. (\ref{equ:gammaInt}) includes implicitly the effects of adsorbate-adsorbate interactions, as both the adsorption energy and the equilibrium coverage depend on such interactions.

To prove Eq.(\ref{equ:gammaInt}), we use the definitions of  $\gamma^{int}$ and 
$E_{ads}$ for a slab of metal in equilibrium with some material X:
\begin{equation}
E_{slab+X} = N E_{bulk} + N_{ads} E_{X} +  2A\gamma^{int}_{hkl},
\label{equ:Eslab_co}
\end{equation}
\begin{equation}
E_{slab+X}=E_{slab}+N_{ads}E_X+N_{ads}E_{ads}.
\label{equ:Eads}
\end{equation}
In the above equations, $E_{slab+X}$ is the total energy of the slab+X system, $E_{X}$ is the total energy per molecule of 
X, and $N_{ads}$ is the number of bonds between slab and X. 
The latter is related to the coverage, $\theta$ and area per surface atom, $A_{at}$, by $\theta = N_{ads}/N_{surf}$ and 
$A=A_{at}N_{surf}$.
Substituting into Eqs. (\ref{equ:Eslab_co}) and (\ref{equ:Eads}) and using Eq. (\ref{equ:Eslab}) yields Eq. 
(\ref{equ:gammaInt}).

For a typical system ($\theta=0.1$, $E_{ads}$=0.5 eV),  
the second term in Eq.(\ref{equ:gammaInt}) is about 0.1 J/m$^2$, or 
10\% of $\gamma_{hkl}$. Change in ratios between various
$\gamma_{hkl}$ will be of the order of 1\%, resulting in very small change in the 
equilibrium shape. This explains the
similarity of nanoparticle shapes observed in a wide variety of
environments: our simulations nicely match experimental
observations, not only for Au clusters
\cite{Cleveland:Structural:1997} but also Au particles on C nanotubes
\cite{Bittencourt:Metallic:2007,Quintana:Light-Induced:2010}, on TiO$_2$
\cite{Sivaramakrishnan:Equilibrium:2010} and on CeO$_2$ \cite{Uchiyama:Systematic:2011}.
\begin{figure}
\includegraphics[width=\columnwidth]{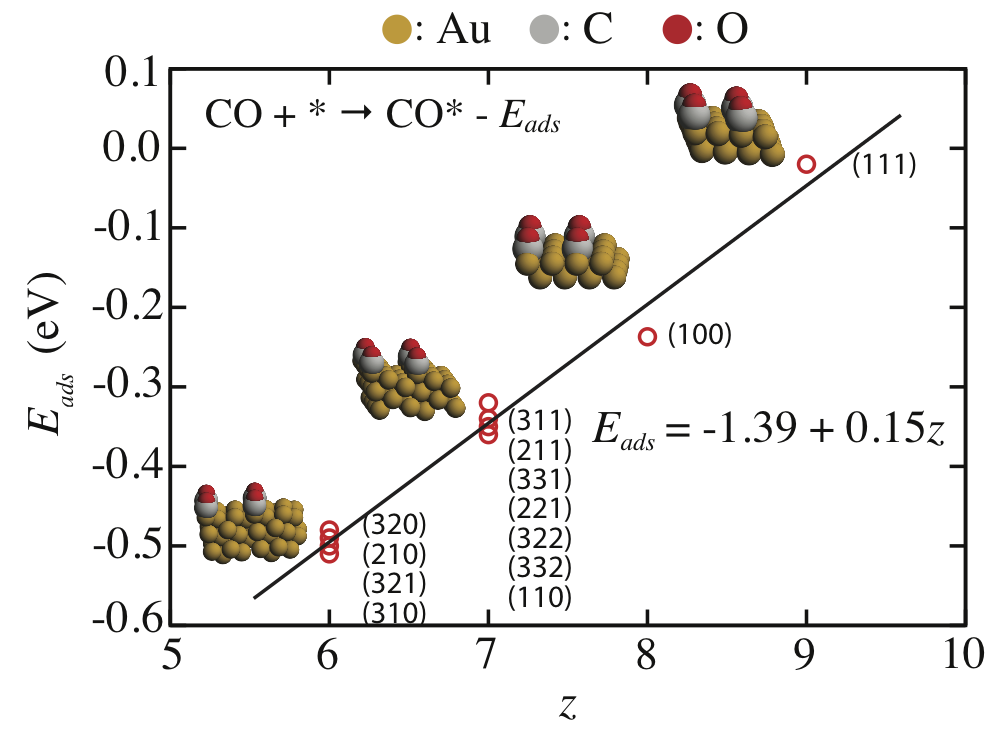}
\caption{(Color online) Adsorption energy of CO on Au($hkl$) as a function of the Au coordination number, $z$, and its linear fit. Some 
characteristic adsorption geometries are shown.}
\label{fig:CO_Eads}
\end{figure}

On the other hand, shape can change dramatically  for very small nanoparticles where bonding on faces might be very
different from bonding on a large surface \cite{ ISI:000257645400007,doi:10.1021/jp981745i} or when small molecules 
with high adsorption energy are adsorbed. The ideal adsorbate to test this idea is CO. 

We calculate the minimum adsorption energy of CO on every
Au($hkl$) with $h,k,l \le 3$.  We consider 
several different adsorption sites to ensure that the global minimum is found; as we are interested in very low CO 
coverage, neighbouring CO molecules maintain a distance of more than 4.2 \AA\ ~at all cases.
In almost every case, CO binds atop the
lowest-coordinated Au atom with adsorption energy being a linear
function of the coordination number of this Au atom, $z$ (Fig.
\ref{fig:CO_Eads})
\cite{Lopez:Catalytic:2002,Remediakis:CO-Oxidation:2005,Remediakis:CO-oxidationGold:2005,Mpourmpakis:Identification:2010}. We use this linear fit  to obtain adsorption energies for the nine $(4kl)$ surfaces. Adsorption energies and adsorption sites are shown in Table \ref{tab:surfaceTensionTable}. 

\begin{figure}
\includegraphics[width=\columnwidth]{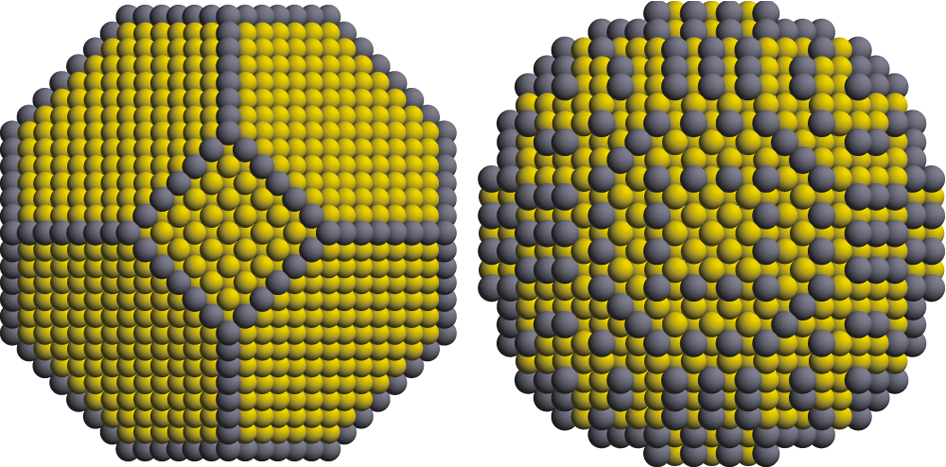}
\caption{(Color online) Left: Model of a typical Au nanoparticle ($d=5.4$ nm, ca. 5000 atoms) in weakly interacting environment (sphericity=93\%, 200 $\mu$mol of active sites per g) Right: a same size nanoparticle in equilibrium with low-pressure CO gas (sphericity=98\%, 400 $\mu$mol of active sites per g). Step and kink atoms are shown in darker color.}
\label{fig:goldNanoparticlesCO}
\end{figure}

We use calculated adsorption energies together with Eq. (\ref{equ:gammaInt}) 
and obtain the equilibrium shape of Au nanoparticles at low CO
coverage shown in Fig. \ref{fig:goldNanoparticlesCO}. 
For rough surfaces, $\gamma_{hkl}$ will be relatively high, but at the
same time $E_{ads}$ will be quite low; this results in a
compensation effect for the two terms in Eq.
(\ref{equ:gammaInt}). As the
different $\gamma^{int}_{hkl}$ are close to each other, the
shape has a much higher sphericity \footnote{Sphericity  equals  $\pi^{1/3}(6V)^{2/3}/A$ where $V$ is the volume and $A$ the area of the nanoparticle; characteristic values are 81\% for a cube, 85\% for an octahedron and 100\% for a sphere.} (98\%) than the shape in
vacuum (93\%), in excellent agreement with experiments
\cite{Ueda:First:2008, Uchiyama:Systematic:2011}. 

Exposure of the nanoparticle to CO gas
makes it much more reactive. This effect has been observed in first-principles simulations of small Au clusters \cite{Remediakis:CO-Oxidation:2005,mckenna07,mckenna09}. We find that the same happens at larger sizes, although it is more prominent for smaller nanoparticles. 
Assuming that all step-edge atoms are
active, the active-site density doubles, increasing from about 200 $\mu$mol/g to 
400 $\mu$mol/g for the nanoparticle shown in Fig. \ref{fig:goldNanoparticlesCO}.

Gold nanoparticles are usually supported on oxides, such as MgO or rutile TiO$_2$.
The interaction between the nanoparticle and the supporting material will also affect its shape. The epitaxial growth will introduce strain in the nanoparticle \cite{Muller:Equilibrium:2000}. More important, the the values of $\gamma_{hkl}$ for the faces attached to the supporting material will be very different. A qualitative picture of this interaction has been presented by Lopez {\em et al.} \cite{lopez04b}.

\section{Summary}

We have developed a method for constructing and characterizing
equilibrium-shaped nanoparticles in thermodynamic equilibrium with their
environment. Using an atomistic version of the Wulff construction, we
generate Cartesian positions of nanoparticles, which can then be used to
analyse structural properties. Our results provide insight into large
nanoparticles that are of interest to catalysis, but are inaccessible by
direct atomistic simulations. The calculated nanoparticles match
experimental results, including the similarity of shapes in weakly
interacting systems as well as the change towards more spherical shapes
upon exposure to reactive gas. The method is easily generalized to other
materials, and might be useful for the improved design of nanomaterials
with tailored physical and chemical properties.

\acknowledgements{
This work was supported by COST action MPMP0901 (NanoTP) and by the Research Council, 
University of Crete. The authors acknowledge support and inspiring discussions with Prof.  S. Farantos.
}

\bibliography{refs}   

\begin{thebibliography}{49}%
\makeatletter
\providecommand \@ifxundefined [1]{%
 \@ifx{#1\undefined}
}%
\providecommand \@ifnum [1]{%
 \ifnum #1\expandafter \@firstoftwo
 \else \expandafter \@secondoftwo
 \fi
}%
\providecommand \@ifx [1]{%
 \ifx #1\expandafter \@firstoftwo
 \else \expandafter \@secondoftwo
 \fi
}%
\providecommand \natexlab [1]{#1}%
\providecommand \enquote  [1]{``#1''}%
\providecommand \bibnamefont  [1]{#1}%
\providecommand \bibfnamefont [1]{#1}%
\providecommand \citenamefont [1]{#1}%
\providecommand \href@noop [0]{\@secondoftwo}%
\providecommand \href [0]{\begingroup \@sanitize@url \@href}%
\providecommand \@href[1]{\@@startlink{#1}\@@href}%
\providecommand \@@href[1]{\endgroup#1\@@endlink}%
\providecommand \@sanitize@url [0]{\catcode `\\12\catcode `\$12\catcode
  `\&12\catcode `\#12\catcode `\^12\catcode `\_12\catcode `\%12\relax}%
\providecommand \@@startlink[1]{}%
\providecommand \@@endlink[0]{}%
\providecommand \url  [0]{\begingroup\@sanitize@url \@url }%
\providecommand \@url [1]{\endgroup\@href {#1}{\urlprefix }}%
\providecommand \urlprefix  [0]{URL }%
\providecommand \Eprint [0]{\href }%
\providecommand \doibase [0]{http://dx.doi.org/}%
\providecommand \selectlanguage [0]{\@gobble}%
\providecommand \bibinfo  [0]{\@secondoftwo}%
\providecommand \bibfield  [0]{\@secondoftwo}%
\providecommand \translation [1]{[#1]}%
\providecommand \BibitemOpen [0]{}%
\providecommand \bibitemStop [0]{}%
\providecommand \bibitemNoStop [0]{.\EOS\space}%
\providecommand \EOS [0]{\spacefactor3000\relax}%
\providecommand \BibitemShut  [1]{\csname bibitem#1\endcsname}%
\let\auto@bib@innerbib\@empty
\bibitem [{\citenamefont {Hammer}\ and\ \citenamefont
  {N{\o}rskov}(1995)}]{Hammer:Why-Gold:1995}%
  \BibitemOpen
  \bibfield  {author} {\bibinfo {author} {\bibfnamefont {B.}~\bibnamefont
  {Hammer}}\ and\ \bibinfo {author} {\bibfnamefont {J.~K.}\ \bibnamefont
  {N{\o}rskov}},\ }\href@noop {} {\bibfield  {journal} {\bibinfo  {journal}
  {Nature}\ }\textbf {\bibinfo {volume} {376}},\ \bibinfo {pages} {238}
  (\bibinfo {year} {1995})}\BibitemShut {NoStop}%
\bibitem [{\citenamefont {Haruta}\ \emph {et~al.}(1989)\citenamefont {Haruta},
  \citenamefont {Yamada}, \citenamefont {Kobayashi},\ and\ \citenamefont
  {Iijima}}]{Haruta:Gold:1989}%
  \BibitemOpen
  \bibfield  {author} {\bibinfo {author} {\bibfnamefont {M.}~\bibnamefont
  {Haruta}}, \bibinfo {author} {\bibfnamefont {N.}~\bibnamefont {Yamada}},
  \bibinfo {author} {\bibfnamefont {T.}~\bibnamefont {Kobayashi}}, \ and\
  \bibinfo {author} {\bibfnamefont {S.}~\bibnamefont {Iijima}},\ }\href@noop {}
  {\bibfield  {journal} {\bibinfo  {journal} {J. Catal.}\ }\textbf {\bibinfo
  {volume} {115}},\ \bibinfo {pages} {301} (\bibinfo {year}
  {1989})}\BibitemShut {NoStop}%
\bibitem [{\citenamefont {Valden}\ \emph {et~al.}(1998)\citenamefont {Valden},
  \citenamefont {Lai},\ and\ \citenamefont {Goodman}}]{Valden:Onset:1998}%
  \BibitemOpen
  \bibfield  {author} {\bibinfo {author} {\bibfnamefont {M.}~\bibnamefont
  {Valden}}, \bibinfo {author} {\bibfnamefont {X.}~\bibnamefont {Lai}}, \ and\
  \bibinfo {author} {\bibfnamefont {D.~W.}\ \bibnamefont {Goodman}},\
  }\href@noop {} {\bibfield  {journal} {\bibinfo  {journal} {Science}\ }\textbf
  {\bibinfo {volume} {281}},\ \bibinfo {pages} {1647} (\bibinfo {year}
  {1998})}\BibitemShut {NoStop}%
\bibitem [{\citenamefont {Uchiyama}\ \emph {et~al.}(2011)\citenamefont
  {Uchiyama}, \citenamefont {Yoshida}, \citenamefont {Kuwauchi}, \citenamefont
  {Ichikawa}, \citenamefont {Shimada}, \citenamefont {Haruta},\ and\
  \citenamefont {Takeda}}]{Uchiyama:Systematic:2011}%
  \BibitemOpen
  \bibfield  {author} {\bibinfo {author} {\bibfnamefont {T.}~\bibnamefont
  {Uchiyama}}, \bibinfo {author} {\bibfnamefont {H.}~\bibnamefont {Yoshida}},
  \bibinfo {author} {\bibfnamefont {Y.}~\bibnamefont {Kuwauchi}}, \bibinfo
  {author} {\bibfnamefont {S.}~\bibnamefont {Ichikawa}}, \bibinfo {author}
  {\bibfnamefont {S.}~\bibnamefont {Shimada}}, \bibinfo {author} {\bibfnamefont
  {M.}~\bibnamefont {Haruta}}, \ and\ \bibinfo {author} {\bibfnamefont
  {S.}~\bibnamefont {Takeda}},\ }\href {\doibase 10.1002/anie.201102487}
  {\bibfield  {journal} {\bibinfo  {journal} {Angew. Chem. Int. Edit.}\
  }\textbf {\bibinfo {volume} {50}},\ \bibinfo {pages} {10157} (\bibinfo {year}
  {2011})}\BibitemShut {NoStop}%
\bibitem [{\citenamefont {Falsig}\ \emph {et~al.}(2008)\citenamefont {Falsig},
  \citenamefont {Hvolbaek}, \citenamefont {Kristensen}, \citenamefont {Jiang},
  \citenamefont {Bligaard}, \citenamefont {Christensen},\ and\ \citenamefont
  {N{\o}rskov}}]{Falsig:Trends:2008}%
  \BibitemOpen
  \bibfield  {author} {\bibinfo {author} {\bibfnamefont {H.}~\bibnamefont
  {Falsig}}, \bibinfo {author} {\bibfnamefont {B.}~\bibnamefont {Hvolbaek}},
  \bibinfo {author} {\bibfnamefont {I.~S.}\ \bibnamefont {Kristensen}},
  \bibinfo {author} {\bibfnamefont {T.}~\bibnamefont {Jiang}}, \bibinfo
  {author} {\bibfnamefont {T.}~\bibnamefont {Bligaard}}, \bibinfo {author}
  {\bibfnamefont {C.~H.}\ \bibnamefont {Christensen}}, \ and\ \bibinfo {author}
  {\bibfnamefont {J.~K.}\ \bibnamefont {N{\o}rskov}},\ }\href@noop {}
  {\bibfield  {journal} {\bibinfo  {journal} {Angew. Chem. Int. Ed.}\ }\textbf
  {\bibinfo {volume} {47}},\ \bibinfo {pages} {4835} (\bibinfo {year}
  {2008})}\BibitemShut {NoStop}%
\bibitem [{\citenamefont {Lopez}\ and\ \citenamefont
  {N{\o}rskov}(2002)}]{Lopez:Catalytic:2002}%
  \BibitemOpen
  \bibfield  {author} {\bibinfo {author} {\bibfnamefont {N.}~\bibnamefont
  {Lopez}}\ and\ \bibinfo {author} {\bibfnamefont {J.~K.}\ \bibnamefont
  {N{\o}rskov}},\ }\href@noop {} {\bibfield  {journal} {\bibinfo  {journal} {J.
  Am. Chem. Soc.}\ }\textbf {\bibinfo {volume} {124}},\ \bibinfo {pages}
  {11262} (\bibinfo {year} {2002})}\BibitemShut {NoStop}%
\bibitem [{\citenamefont {Remediakis}\ \emph
  {et~al.}(2005{\natexlab{a}})\citenamefont {Remediakis}, \citenamefont
  {Lopez},\ and\ \citenamefont {N{\o}rskov}}]{Remediakis:CO-Oxidation:2005}%
  \BibitemOpen
  \bibfield  {author} {\bibinfo {author} {\bibfnamefont {I.~N.}\ \bibnamefont
  {Remediakis}}, \bibinfo {author} {\bibfnamefont {N.}~\bibnamefont {Lopez}}, \
  and\ \bibinfo {author} {\bibfnamefont {J.~K.}\ \bibnamefont {N{\o}rskov}},\
  }\href@noop {} {\bibfield  {journal} {\bibinfo  {journal} {Angew. Chem. Int.
  Edit.}\ }\textbf {\bibinfo {volume} {44}},\ \bibinfo {pages} {1824} (\bibinfo
  {year} {2005}{\natexlab{a}})}\BibitemShut {NoStop}%
\bibitem [{\citenamefont {Remediakis}\ \emph
  {et~al.}(2005{\natexlab{b}})\citenamefont {Remediakis}, \citenamefont
  {Lopez},\ and\ \citenamefont
  {N{\o}rskov}}]{Remediakis:CO-oxidationGold:2005}%
  \BibitemOpen
  \bibfield  {author} {\bibinfo {author} {\bibfnamefont {I.~N.}\ \bibnamefont
  {Remediakis}}, \bibinfo {author} {\bibfnamefont {N.}~\bibnamefont {Lopez}}, \
  and\ \bibinfo {author} {\bibfnamefont {J.~K.}\ \bibnamefont {N{\o}rskov}},\
  }\href@noop {} {\bibfield  {journal} {\bibinfo  {journal} {Appl. Catal.
  A-Gen.}\ }\textbf {\bibinfo {volume} {291}},\ \bibinfo {pages} {13} (\bibinfo
  {year} {2005}{\natexlab{b}})}\BibitemShut {NoStop}%
\bibitem [{\citenamefont {Kim}\ \emph {et~al.}(2001)\citenamefont {Kim},
  \citenamefont {Johnson},\ and\ \citenamefont {Hupp}}]{ISI:000173708300001}%
  \BibitemOpen
  \bibfield  {author} {\bibinfo {author} {\bibfnamefont {Y.}~\bibnamefont
  {Kim}}, \bibinfo {author} {\bibfnamefont {R.}~\bibnamefont {Johnson}}, \ and\
  \bibinfo {author} {\bibfnamefont {J.}~\bibnamefont {Hupp}},\ }\href@noop {}
  {\bibfield  {journal} {\bibinfo  {journal} {{Nano Lett.}}\ }\textbf {\bibinfo
  {volume} {{1}}},\ \bibinfo {pages} {{165}} (\bibinfo {year}
  {{2001}})}\BibitemShut {NoStop}%
\bibitem [{\citenamefont {Boisselier}\ and\ \citenamefont
  {Astruc}(2009)}]{Boisselier:Gold:2009}%
  \BibitemOpen
  \bibfield  {author} {\bibinfo {author} {\bibfnamefont {E.}~\bibnamefont
  {Boisselier}}\ and\ \bibinfo {author} {\bibfnamefont {D.}~\bibnamefont
  {Astruc}},\ }\href@noop {} {\bibfield  {journal} {\bibinfo  {journal} {Chem.
  Soc. Rev.}\ }\textbf {\bibinfo {volume} {38}},\ \bibinfo {pages} {1759}
  (\bibinfo {year} {2009})}\BibitemShut {NoStop}%
\bibitem [{\citenamefont {Lal}\ \emph {et~al.}(2007)\citenamefont {Lal},
  \citenamefont {Link},\ and\ \citenamefont {Halas}}]{Lal:Nano-optics:2007}%
  \BibitemOpen
  \bibfield  {author} {\bibinfo {author} {\bibfnamefont {S.}~\bibnamefont
  {Lal}}, \bibinfo {author} {\bibfnamefont {S.}~\bibnamefont {Link}}, \ and\
  \bibinfo {author} {\bibfnamefont {N.~J.}\ \bibnamefont {Halas}},\ }\href@noop
  {} {\bibfield  {journal} {\bibinfo  {journal} {Nat. Photonics}\ }\textbf
  {\bibinfo {volume} {1}},\ \bibinfo {pages} {641} (\bibinfo {year}
  {2007})}\BibitemShut {NoStop}%
\bibitem [{\citenamefont {Lu}\ \emph {et~al.}(2002)\citenamefont {Lu},
  \citenamefont {Yin}, \citenamefont {Li},\ and\ \citenamefont
  {Xia}}]{Lu:Synthesis:2002}%
  \BibitemOpen
  \bibfield  {author} {\bibinfo {author} {\bibfnamefont {Y.}~\bibnamefont
  {Lu}}, \bibinfo {author} {\bibfnamefont {Y.}~\bibnamefont {Yin}}, \bibinfo
  {author} {\bibfnamefont {Z.}~\bibnamefont {Li}}, \ and\ \bibinfo {author}
  {\bibfnamefont {Y.}~\bibnamefont {Xia}},\ }\href@noop {} {\bibfield
  {journal} {\bibinfo  {journal} {Nano Lett.}\ }\textbf {\bibinfo {volume}
  {2}},\ \bibinfo {pages} {785} (\bibinfo {year} {2002})}\BibitemShut {NoStop}%
\bibitem [{Note1()}]{Note1}%
  \BibitemOpen
  \bibinfo {note} {For cube of edge $a$, $\Delta E={\begingroup {\mathchar
  '26\mkern -9muh}^2\pi ^2\endgroup \over 2ma^2}$; for a sphere of radius $r$,
  $\Delta E={\begingroup {\mathchar '26\mkern -9muh}^2(\chi _{11}-\chi
  _{10})\endgroup \over 2mr^2}$, where $\chi _{11},\chi _{10}$ are the first
  two roots of the spherical Bessel function $j_1(x)$}\BibitemShut {NoStop}%
\bibitem [{\citenamefont {Wen}\ and\ \citenamefont
  {Zhang}(2007)}]{Wen:Surface:2007}%
  \BibitemOpen
  \bibfield  {author} {\bibinfo {author} {\bibfnamefont {Y.-N.}\ \bibnamefont
  {Wen}}\ and\ \bibinfo {author} {\bibfnamefont {J.-M.}\ \bibnamefont
  {Zhang}},\ }\href@noop {} {\bibfield  {journal} {\bibinfo  {journal} {Solid
  State Commun.}\ }\textbf {\bibinfo {volume} {144}},\ \bibinfo {pages} {163}
  (\bibinfo {year} {2007})}\BibitemShut {NoStop}%
\bibitem [{\citenamefont {Vitos}\ \emph {et~al.}(1998)\citenamefont {Vitos},
  \citenamefont {Ruban}, \citenamefont {Skriver},\ and\ \citenamefont
  {Kollαr}}]{Vitos:The-surface:1998}%
  \BibitemOpen
  \bibfield  {author} {\bibinfo {author} {\bibfnamefont {L.}~\bibnamefont
  {Vitos}}, \bibinfo {author} {\bibfnamefont {A.~V.}\ \bibnamefont {Ruban}},
  \bibinfo {author} {\bibfnamefont {H.~L.}\ \bibnamefont {Skriver}}, \ and\
  \bibinfo {author} {\bibfnamefont {J.}~\bibnamefont {Kollαr}},\ }\href@noop
  {} {\bibfield  {journal} {\bibinfo  {journal} {Surf. Sci.}\ }\textbf
  {\bibinfo {volume} {411}},\ \bibinfo {pages} {186} (\bibinfo {year}
  {1998})}\BibitemShut {NoStop}%
\bibitem [{\citenamefont {Galanakis}\ \emph {et~al.}(2002)\citenamefont
  {Galanakis}, \citenamefont {Bihlmayer}, \citenamefont {Bellini},
  \citenamefont {Papanikolaou}, \citenamefont {Zeller}, \citenamefont
  {Bl{\"u}gel},\ and\ \citenamefont {Dederichs}}]{Galanakis:Broken-bond:2002}%
  \BibitemOpen
  \bibfield  {author} {\bibinfo {author} {\bibfnamefont {I.}~\bibnamefont
  {Galanakis}}, \bibinfo {author} {\bibfnamefont {G.}~\bibnamefont
  {Bihlmayer}}, \bibinfo {author} {\bibfnamefont {V.}~\bibnamefont {Bellini}},
  \bibinfo {author} {\bibfnamefont {N.}~\bibnamefont {Papanikolaou}}, \bibinfo
  {author} {\bibfnamefont {R.}~\bibnamefont {Zeller}}, \bibinfo {author}
  {\bibfnamefont {S.}~\bibnamefont {Bl{\"u}gel}}, \ and\ \bibinfo {author}
  {\bibfnamefont {P.~H.}\ \bibnamefont {Dederichs}},\ }\href@noop {} {\bibfield
   {journal} {\bibinfo  {journal} {Europhys. Lett.}\ }\textbf {\bibinfo
  {volume} {58}},\ \bibinfo {pages} {751} (\bibinfo {year} {2002})}\BibitemShut
  {NoStop}%
\bibitem [{\citenamefont {Ueda}\ \emph {et~al.}(2008)\citenamefont {Ueda},
  \citenamefont {Kawasaki}, \citenamefont {Hasegawa}, \citenamefont {Tanji},\
  and\ \citenamefont {Ichihashi}}]{Ueda:First:2008}%
  \BibitemOpen
  \bibfield  {author} {\bibinfo {author} {\bibfnamefont {K.}~\bibnamefont
  {Ueda}}, \bibinfo {author} {\bibfnamefont {T.}~\bibnamefont {Kawasaki}},
  \bibinfo {author} {\bibfnamefont {H.}~\bibnamefont {Hasegawa}}, \bibinfo
  {author} {\bibfnamefont {T.}~\bibnamefont {Tanji}}, \ and\ \bibinfo {author}
  {\bibfnamefont {M.}~\bibnamefont {Ichihashi}},\ }\href@noop {} {\bibfield
  {journal} {\bibinfo  {journal} {Surf. Interface Anal.}\ }\textbf {\bibinfo
  {volume} {40}},\ \bibinfo {pages} {1725} (\bibinfo {year}
  {2008})}\BibitemShut {NoStop}%
\bibitem [{\citenamefont {McKenna}(2009)}]{mckenna09}%
  \BibitemOpen
  \bibfield  {author} {\bibinfo {author} {\bibfnamefont {K.~P.}\ \bibnamefont
  {McKenna}},\ }\href {\doibase {10.1039/b821408p}} {\bibfield  {journal}
  {\bibinfo  {journal} {{Phys. Chem. Chem. Phys.}}\ }\textbf {\bibinfo {volume}
  {{11}}},\ \bibinfo {pages} {4145} (\bibinfo {year} {2009})}\BibitemShut
  {NoStop}%
\bibitem [{\citenamefont {Clausen}\ \emph {et~al.}(1994)\citenamefont
  {Clausen}, \citenamefont {Schi{\o}tz}, \citenamefont {Gr{\aa}b{\ae}k},
  \citenamefont {Ovesen}, \citenamefont {Jacobsen}, \citenamefont
  {N{\o}rskov},\ and\ \citenamefont {Tops{\o}e}}]{Clausen:Wetting:1994}%
  \BibitemOpen
  \bibfield  {author} {\bibinfo {author} {\bibfnamefont {B.~S.}\ \bibnamefont
  {Clausen}}, \bibinfo {author} {\bibfnamefont {J.}~\bibnamefont {Schi{\o}tz}},
  \bibinfo {author} {\bibfnamefont {L.}~\bibnamefont {Gr{\aa}b{\ae}k}},
  \bibinfo {author} {\bibfnamefont {C.~V.}\ \bibnamefont {Ovesen}}, \bibinfo
  {author} {\bibfnamefont {K.~W.}\ \bibnamefont {Jacobsen}}, \bibinfo {author}
  {\bibfnamefont {J.~K.}\ \bibnamefont {N{\o}rskov}}, \ and\ \bibinfo {author}
  {\bibfnamefont {H.}~\bibnamefont {Tops{\o}e}},\ }\href@noop {} {\bibfield
  {journal} {\bibinfo  {journal} {Top. Catal.}\ }\textbf {\bibinfo {volume}
  {1}},\ \bibinfo {pages} {367} (\bibinfo {year} {1994})}\BibitemShut {NoStop}%
\bibitem [{\citenamefont {Hansen}\ \emph {et~al.}(2002)\citenamefont {Hansen},
  \citenamefont {Wagner}, \citenamefont {Helveg}, \citenamefont
  {Rostrup-Nielsen}, \citenamefont {BS},\ and\ \citenamefont
  {Tops{\o}e}}]{Hansen:Atom-resolved:2002}%
  \BibitemOpen
  \bibfield  {author} {\bibinfo {author} {\bibfnamefont {P.~L.}\ \bibnamefont
  {Hansen}}, \bibinfo {author} {\bibfnamefont {J.~B.}\ \bibnamefont {Wagner}},
  \bibinfo {author} {\bibfnamefont {S.}~\bibnamefont {Helveg}}, \bibinfo
  {author} {\bibfnamefont {J.~R.}\ \bibnamefont {Rostrup-Nielsen}}, \bibinfo
  {author} {\bibfnamefont {B.~S.~C.}\ \bibnamefont {BS}}, \ and\ \bibinfo
  {author} {\bibfnamefont {H.}~\bibnamefont {Tops{\o}e}},\ }\href@noop {}
  {\bibfield  {journal} {\bibinfo  {journal} {Science}\ }\textbf {\bibinfo
  {volume} {295}},\ \bibinfo {pages} {2053} (\bibinfo {year}
  {2002})}\BibitemShut {NoStop}%
\bibitem [{\citenamefont {M{\"u}ller}\ and\ \citenamefont
  {Kern}(2000)}]{Muller:Equilibrium:2000}%
  \BibitemOpen
  \bibfield  {author} {\bibinfo {author} {\bibfnamefont {P.}~\bibnamefont
  {M{\"u}ller}}\ and\ \bibinfo {author} {\bibfnamefont {R.}~\bibnamefont
  {Kern}},\ }\href@noop {} {\bibfield  {journal} {\bibinfo  {journal} {Surf.
  Sci.}\ }\textbf {\bibinfo {volume} {457}},\ \bibinfo {pages} {229} (\bibinfo
  {year} {2000})}\BibitemShut {NoStop}%
\bibitem [{\citenamefont {Barnard}\ and\ \citenamefont
  {Zapol}(2004)}]{barnard04}%
  \BibitemOpen
  \bibfield  {author} {\bibinfo {author} {\bibfnamefont {A.}~\bibnamefont
  {Barnard}}\ and\ \bibinfo {author} {\bibfnamefont {P.}~\bibnamefont
  {Zapol}},\ }\href {\doibase {10.1063/1.1775770}} {\bibfield  {journal}
  {\bibinfo  {journal} {{J. Chem. Phys.}}\ }\textbf {\bibinfo {volume} {121}},\
  \bibinfo {pages} {4276} (\bibinfo {year} {2004})}\BibitemShut {NoStop}%
\bibitem [{\citenamefont {Barnard}\ and\ \citenamefont
  {Curtiss}(2005)}]{barnard05}%
  \BibitemOpen
  \bibfield  {author} {\bibinfo {author} {\bibfnamefont {A.~S.}\ \bibnamefont
  {Barnard}}\ and\ \bibinfo {author} {\bibfnamefont {L.~A.}\ \bibnamefont
  {Curtiss}},\ }\href {\doibase 10.1021/nl050355m} {\bibfield  {journal}
  {\bibinfo  {journal} {Nano Lett.}\ }\textbf {\bibinfo {volume} {5}},\
  \bibinfo {pages} {1261} (\bibinfo {year} {2005})}\BibitemShut {NoStop}%
\bibitem [{\citenamefont {Hadjisavvas}\ \emph {et~al.}(2006)\citenamefont
  {Hadjisavvas}, \citenamefont {Remediakis},\ and\ \citenamefont
  {Kelires}}]{Hadjisavvas:Insights:2006}%
  \BibitemOpen
  \bibfield  {author} {\bibinfo {author} {\bibfnamefont {G.}~\bibnamefont
  {Hadjisavvas}}, \bibinfo {author} {\bibfnamefont {I.~N.}\ \bibnamefont
  {Remediakis}}, \ and\ \bibinfo {author} {\bibfnamefont {P.~C.}\ \bibnamefont
  {Kelires}},\ }\href@noop {} {\bibfield  {journal} {\bibinfo  {journal} {Phys.
  Rev. B}\ }\textbf {\bibinfo {volume} {74}},\ \bibinfo {pages} {165419}
  (\bibinfo {year} {2006})}\BibitemShut {NoStop}%
\bibitem [{\citenamefont {Kopidakis}\ \emph {et~al.}(2007)\citenamefont
  {Kopidakis}, \citenamefont {Remediakis}, \citenamefont {Fyta},\ and\
  \citenamefont {Kelires}}]{Kopidakis:Atomic:2007}%
  \BibitemOpen
  \bibfield  {author} {\bibinfo {author} {\bibfnamefont {G.}~\bibnamefont
  {Kopidakis}}, \bibinfo {author} {\bibfnamefont {I.~N.}\ \bibnamefont
  {Remediakis}}, \bibinfo {author} {\bibfnamefont {M.~G.}\ \bibnamefont
  {Fyta}}, \ and\ \bibinfo {author} {\bibfnamefont {P.~C.}\ \bibnamefont
  {Kelires}},\ }\href@noop {} {\bibfield  {journal} {\bibinfo  {journal} {Diam.
  Relat. Mater.}\ }\textbf {\bibinfo {volume} {16}},\ \bibinfo {pages} {1875}
  (\bibinfo {year} {2007})}\BibitemShut {NoStop}%
\bibitem [{\citenamefont {Mittendorfer}\ \emph {et~al.}(2007)\citenamefont
  {Mittendorfer}, \citenamefont {Seriani}, \citenamefont {Dubay},\ and\
  \citenamefont {Kresse}}]{mittendorfer07}%
  \BibitemOpen
  \bibfield  {author} {\bibinfo {author} {\bibfnamefont {F.}~\bibnamefont
  {Mittendorfer}}, \bibinfo {author} {\bibfnamefont {N.}~\bibnamefont
  {Seriani}}, \bibinfo {author} {\bibfnamefont {O.}~\bibnamefont {Dubay}}, \
  and\ \bibinfo {author} {\bibfnamefont {G.}~\bibnamefont {Kresse}},\ }\href
  {\doibase {10.1103/PhysRevB.76.233413}} {\bibfield  {journal} {\bibinfo
  {journal} {{Phys. Rev. B}}\ }\textbf {\bibinfo {volume} {{76}}},\ \bibinfo
  {pages} {233413} (\bibinfo {year} {{2007}})}\BibitemShut {NoStop}%
\bibitem [{\citenamefont {Soon}\ \emph {et~al.}(2008)\citenamefont {Soon},
  \citenamefont {Wong}, \citenamefont {Delley},\ and\ \citenamefont
  {Stampfl}}]{Soon:Morphology:2008}%
  \BibitemOpen
  \bibfield  {author} {\bibinfo {author} {\bibfnamefont {A.}~\bibnamefont
  {Soon}}, \bibinfo {author} {\bibfnamefont {L.}~\bibnamefont {Wong}}, \bibinfo
  {author} {\bibfnamefont {B.}~\bibnamefont {Delley}}, \ and\ \bibinfo {author}
  {\bibfnamefont {C.}~\bibnamefont {Stampfl}},\ }\href@noop {} {\bibfield
  {journal} {\bibinfo  {journal} {Phys. Rev. B}\ }\textbf {\bibinfo {volume}
  {77}},\ \bibinfo {pages} {125423} (\bibinfo {year} {2008})}\BibitemShut
  {NoStop}%
\bibitem [{\citenamefont {Shi}\ and\ \citenamefont
  {Stampfl}(2008)}]{Shi:Shape:2008}%
  \BibitemOpen
  \bibfield  {author} {\bibinfo {author} {\bibfnamefont {H.}~\bibnamefont
  {Shi}}\ and\ \bibinfo {author} {\bibfnamefont {C.}~\bibnamefont {Stampfl}},\
  }\href@noop {} {\bibfield  {journal} {\bibinfo  {journal} {Phys. Rev. B}\
  }\textbf {\bibinfo {volume} {77}},\ \bibinfo {pages} {094127} (\bibinfo
  {year} {2008})}\BibitemShut {NoStop}%
\bibitem [{\citenamefont {Honkala}\ \emph {et~al.}(2005)\citenamefont
  {Honkala}, \citenamefont {Hellman}, \citenamefont {Remediakis}, \citenamefont
  {Logadottir}, \citenamefont {Carlsson}, \citenamefont {Dahl}, \citenamefont
  {Christensen},\ and\ \citenamefont {N{\o}rskov}}]{Honkala:Ammonia:2005}%
  \BibitemOpen
  \bibfield  {author} {\bibinfo {author} {\bibfnamefont {K.}~\bibnamefont
  {Honkala}}, \bibinfo {author} {\bibfnamefont {A.}~\bibnamefont {Hellman}},
  \bibinfo {author} {\bibfnamefont {I.~N.}\ \bibnamefont {Remediakis}},
  \bibinfo {author} {\bibfnamefont {A.}~\bibnamefont {Logadottir}}, \bibinfo
  {author} {\bibfnamefont {A.}~\bibnamefont {Carlsson}}, \bibinfo {author}
  {\bibfnamefont {S.}~\bibnamefont {Dahl}}, \bibinfo {author} {\bibfnamefont
  {C.}~\bibnamefont {Christensen}}, \ and\ \bibinfo {author} {\bibfnamefont
  {J.~K.}\ \bibnamefont {N{\o}rskov}},\ }\href@noop {} {\bibfield  {journal}
  {\bibinfo  {journal} {Science}\ }\textbf {\bibinfo {volume} {307}},\ \bibinfo
  {pages} {558} (\bibinfo {year} {2005})}\BibitemShut {NoStop}%
\bibitem [{\citenamefont {Hellman}\ \emph {et~al.}(2006)\citenamefont
  {Hellman}, \citenamefont {Honkala}, \citenamefont {Remediakis}, \citenamefont
  {Logadottir}, \citenamefont {Carlsson}, \citenamefont {Dahl}, \citenamefont
  {Christensen},\ and\ \citenamefont {N{\o}rskov}}]{hellman06}%
  \BibitemOpen
  \bibfield  {author} {\bibinfo {author} {\bibfnamefont {A.}~\bibnamefont
  {Hellman}}, \bibinfo {author} {\bibfnamefont {K.}~\bibnamefont {Honkala}},
  \bibinfo {author} {\bibfnamefont {I.~N.}\ \bibnamefont {Remediakis}},
  \bibinfo {author} {\bibfnamefont {A.}~\bibnamefont {Logadottir}}, \bibinfo
  {author} {\bibfnamefont {A.}~\bibnamefont {Carlsson}}, \bibinfo {author}
  {\bibfnamefont {S.}~\bibnamefont {Dahl}}, \bibinfo {author} {\bibfnamefont
  {C.~H.}\ \bibnamefont {Christensen}}, \ and\ \bibinfo {author} {\bibfnamefont
  {J.~K.}\ \bibnamefont {N{\o}rskov}},\ }\href {\doibase
  {10.1016/j.susc.2005.11.070}} {\bibfield  {journal} {\bibinfo  {journal}
  {{Surf. Sci.}}\ }\textbf {\bibinfo {volume} {600}},\ \bibinfo {pages} {4264}
  (\bibinfo {year} {2006})}\BibitemShut {NoStop}%
\bibitem [{\citenamefont {Hellman}\ \emph {et~al.}(2009)\citenamefont
  {Hellman}, \citenamefont {Honkala}, \citenamefont {Remediakis}, \citenamefont
  {Logadottir}, \citenamefont {Carlsson}, \citenamefont {Dahl}, \citenamefont
  {Christensen},\ and\ \citenamefont {N{\o}rskov}}]{hellman09}%
  \BibitemOpen
  \bibfield  {author} {\bibinfo {author} {\bibfnamefont {A.}~\bibnamefont
  {Hellman}}, \bibinfo {author} {\bibfnamefont {K.}~\bibnamefont {Honkala}},
  \bibinfo {author} {\bibfnamefont {I.~N.}\ \bibnamefont {Remediakis}},
  \bibinfo {author} {\bibfnamefont {A.}~\bibnamefont {Logadottir}}, \bibinfo
  {author} {\bibfnamefont {A.}~\bibnamefont {Carlsson}}, \bibinfo {author}
  {\bibfnamefont {S.}~\bibnamefont {Dahl}}, \bibinfo {author} {\bibfnamefont
  {C.~H.}\ \bibnamefont {Christensen}}, \ and\ \bibinfo {author} {\bibfnamefont
  {J.~K.}\ \bibnamefont {N{\o}rskov}},\ }\href {\doibase
  {10.1016/j.susc.2008.10.059}} {\bibfield  {journal} {\bibinfo  {journal}
  {Surf. Sci.}\ }\textbf {\bibinfo {volume} {603}},\ \bibinfo {pages} {1731}
  (\bibinfo {year} {2009})}\BibitemShut {NoStop}%
\bibitem [{\citenamefont {Herring}(1951)}]{Herring:Some:1951}%
  \BibitemOpen
  \bibfield  {author} {\bibinfo {author} {\bibfnamefont {C.}~\bibnamefont
  {Herring}},\ }\href {\doibase 10.1103/PhysRev.82.87} {\bibfield  {journal}
  {\bibinfo  {journal} {Phys. Rev.}\ }\textbf {\bibinfo {volume} {82}},\
  \bibinfo {pages} {87} (\bibinfo {year} {1951})}\BibitemShut {NoStop}%
\bibitem [{\citenamefont {Roosen}\ \emph {et~al.}(1998)\citenamefont {Roosen},
  \citenamefont {McCormack},\ and\ \citenamefont
  {Carter}}]{Roosen:Wulffman::1998}%
  \BibitemOpen
  \bibfield  {author} {\bibinfo {author} {\bibfnamefont {A.~R.}\ \bibnamefont
  {Roosen}}, \bibinfo {author} {\bibfnamefont {R.~P.}\ \bibnamefont
  {McCormack}}, \ and\ \bibinfo {author} {\bibfnamefont {W.~C.}\ \bibnamefont
  {Carter}},\ }\href@noop {} {\bibfield  {journal} {\bibinfo  {journal} {Comp.
  Mater. Sci.}\ }\textbf {\bibinfo {volume} {11}},\ \bibinfo {pages} {16}
  (\bibinfo {year} {1998})}\BibitemShut {NoStop}%
\bibitem [{\citenamefont {Cleveland}\ \emph {et~al.}(1997)\citenamefont
  {Cleveland}, \citenamefont {Landman}, \citenamefont {Shafigullin},
  \citenamefont {Stephens},\ and\ \citenamefont
  {Whetten}}]{Cleveland:Structural:1997}%
  \BibitemOpen
  \bibfield  {author} {\bibinfo {author} {\bibfnamefont {C.~L.}\ \bibnamefont
  {Cleveland}}, \bibinfo {author} {\bibfnamefont {U.}~\bibnamefont {Landman}},
  \bibinfo {author} {\bibfnamefont {M.~N.}\ \bibnamefont {Shafigullin}},
  \bibinfo {author} {\bibfnamefont {P.~W.}\ \bibnamefont {Stephens}}, \ and\
  \bibinfo {author} {\bibfnamefont {R.~L.}\ \bibnamefont {Whetten}},\
  }\href@noop {} {\bibfield  {journal} {\bibinfo  {journal} {Zeit. Phys. D}\
  }\textbf {\bibinfo {volume} {40}},\ \bibinfo {pages} {503} (\bibinfo {year}
  {1997})}\BibitemShut {NoStop}%
\bibitem [{\citenamefont {Baletto}\ and\ \citenamefont
  {Ferrando}(2005)}]{baletto05}%
  \BibitemOpen
  \bibfield  {author} {\bibinfo {author} {\bibfnamefont {F.}~\bibnamefont
  {Baletto}}\ and\ \bibinfo {author} {\bibfnamefont {R.}~\bibnamefont
  {Ferrando}},\ }\href {\doibase 10.1103/RevModPhys.77.371} {\bibfield
  {journal} {\bibinfo  {journal} {Rev. Mod. Phys.}\ }\textbf {\bibinfo {volume}
  {77}},\ \bibinfo {pages} {371} (\bibinfo {year} {2005})}\BibitemShut
  {NoStop}%
\bibitem [{\citenamefont {Vanderbilt}(1990)}]{vanderbilt90}%
  \BibitemOpen
  \bibfield  {author} {\bibinfo {author} {\bibfnamefont {D.}~\bibnamefont
  {Vanderbilt}},\ }\href {\doibase 10.1103/PhysRevB.41.7892} {\bibfield
  {journal} {\bibinfo  {journal} {Phys. Rev. B}\ }\textbf {\bibinfo {volume}
  {41}},\ \bibinfo {pages} {7892} (\bibinfo {year} {1990})}\BibitemShut
  {NoStop}%
\bibitem [{\citenamefont {Hammer}\ \emph {et~al.}(1999)\citenamefont {Hammer},
  \citenamefont {Hansen},\ and\ \citenamefont
  {N\o{}rskov}}]{Hammer:Improved:1999}%
  \BibitemOpen
  \bibfield  {author} {\bibinfo {author} {\bibfnamefont {B.}~\bibnamefont
  {Hammer}}, \bibinfo {author} {\bibfnamefont {L.~B.}\ \bibnamefont {Hansen}},
  \ and\ \bibinfo {author} {\bibfnamefont {J.~K.}\ \bibnamefont {N\o{}rskov}},\
  }\href {\doibase 10.1103/PhysRevB.59.7413} {\bibfield  {journal} {\bibinfo
  {journal} {Phys. Rev. B}\ }\textbf {\bibinfo {volume} {59}},\ \bibinfo
  {pages} {7413} (\bibinfo {year} {1999})}\BibitemShut {NoStop}%
\bibitem [{\citenamefont {Fuchs}\ \emph {et~al.}(1998)\citenamefont {Fuchs},
  \citenamefont {Bockstedte}, \citenamefont {Pehlke},\ and\ \citenamefont
  {Scheffler}}]{fuchs98}%
  \BibitemOpen
  \bibfield  {author} {\bibinfo {author} {\bibfnamefont {M.}~\bibnamefont
  {Fuchs}}, \bibinfo {author} {\bibfnamefont {M.}~\bibnamefont {Bockstedte}},
  \bibinfo {author} {\bibfnamefont {E.}~\bibnamefont {Pehlke}}, \ and\ \bibinfo
  {author} {\bibfnamefont {M.}~\bibnamefont {Scheffler}},\ }\href {\doibase
  10.1103/PhysRevB.57.2134} {\bibfield  {journal} {\bibinfo  {journal} {Phys.
  Rev. B}\ }\textbf {\bibinfo {volume} {57}},\ \bibinfo {pages} {2134}
  (\bibinfo {year} {1998})}\BibitemShut {NoStop}%
\bibitem [{\citenamefont {Lopez}\ \emph
  {et~al.}(2004{\natexlab{a}})\citenamefont {Lopez}, \citenamefont {Janssens},
  \citenamefont {Clausen}, \citenamefont {Xu}, \citenamefont {Mavrikakis},
  \citenamefont {Bligaard},\ and\ \citenamefont
  {N{\o}rskov}}]{Lopez:On-the-origin:2004}%
  \BibitemOpen
  \bibfield  {author} {\bibinfo {author} {\bibfnamefont {N.}~\bibnamefont
  {Lopez}}, \bibinfo {author} {\bibfnamefont {T.}~\bibnamefont {Janssens}},
  \bibinfo {author} {\bibfnamefont {B.}~\bibnamefont {Clausen}}, \bibinfo
  {author} {\bibfnamefont {Y.}~\bibnamefont {Xu}}, \bibinfo {author}
  {\bibfnamefont {M.}~\bibnamefont {Mavrikakis}}, \bibinfo {author}
  {\bibfnamefont {T.}~\bibnamefont {Bligaard}}, \ and\ \bibinfo {author}
  {\bibfnamefont {J.~K.}\ \bibnamefont {N{\o}rskov}},\ }\href@noop {}
  {\bibfield  {journal} {\bibinfo  {journal} {J. Catal.}\ }\textbf {\bibinfo
  {volume} {223}},\ \bibinfo {pages} {232} (\bibinfo {year}
  {2004}{\natexlab{a}})}\BibitemShut {NoStop}%
\bibitem [{\citenamefont {Galanis}\ \emph {et~al.}(2010)\citenamefont
  {Galanis}, \citenamefont {Remediakis},\ and\ \citenamefont
  {Kopidakis}}]{Galanis:Mechanical:2010}%
  \BibitemOpen
  \bibfield  {author} {\bibinfo {author} {\bibfnamefont {N.~V.}\ \bibnamefont
  {Galanis}}, \bibinfo {author} {\bibfnamefont {I.~N.}\ \bibnamefont
  {Remediakis}}, \ and\ \bibinfo {author} {\bibfnamefont {G.}~\bibnamefont
  {Kopidakis}},\ }\href@noop {} {\bibfield  {journal} {\bibinfo  {journal}
  {Phys. Status Solidi C}\ }\textbf {\bibinfo {volume} {7}},\ \bibinfo {pages}
  {1372} (\bibinfo {year} {2010})}\BibitemShut {NoStop}%
\bibitem [{\citenamefont {Bittencourt}\ \emph {et~al.}(2007)\citenamefont
  {Bittencourt}, \citenamefont {Felten}, \citenamefont {Douhard}, \citenamefont
  {Colomer}, \citenamefont {Tendeloo}, \citenamefont {Drube}, \citenamefont
  {Ghijsen},\ and\ \citenamefont {Pireaux}}]{Bittencourt:Metallic:2007}%
  \BibitemOpen
  \bibfield  {author} {\bibinfo {author} {\bibfnamefont {C.}~\bibnamefont
  {Bittencourt}}, \bibinfo {author} {\bibfnamefont {A.}~\bibnamefont {Felten}},
  \bibinfo {author} {\bibfnamefont {B.}~\bibnamefont {Douhard}}, \bibinfo
  {author} {\bibfnamefont {J.-F.}\ \bibnamefont {Colomer}}, \bibinfo {author}
  {\bibfnamefont {G.~V.}\ \bibnamefont {Tendeloo}}, \bibinfo {author}
  {\bibfnamefont {W.}~\bibnamefont {Drube}}, \bibinfo {author} {\bibfnamefont
  {J.}~\bibnamefont {Ghijsen}}, \ and\ \bibinfo {author} {\bibfnamefont
  {J.-J.}\ \bibnamefont {Pireaux}},\ }\href@noop {} {\bibfield  {journal}
  {\bibinfo  {journal} {Surf. Sci.}\ }\textbf {\bibinfo {volume} {601}},\
  \bibinfo {pages} {2800} (\bibinfo {year} {2007})}\BibitemShut {NoStop}%
\bibitem [{\citenamefont {Quintana}\ \emph {et~al.}(2010)\citenamefont
  {Quintana}, \citenamefont {Ke}, \citenamefont {Tendeloo}, \citenamefont
  {Meneghetti}, \citenamefont {Bittencourt},\ and\ \citenamefont
  {Prato}}]{Quintana:Light-Induced:2010}%
  \BibitemOpen
  \bibfield  {author} {\bibinfo {author} {\bibfnamefont {M.}~\bibnamefont
  {Quintana}}, \bibinfo {author} {\bibfnamefont {X.}~\bibnamefont {Ke}},
  \bibinfo {author} {\bibfnamefont {G.~V.}\ \bibnamefont {Tendeloo}}, \bibinfo
  {author} {\bibfnamefont {M.}~\bibnamefont {Meneghetti}}, \bibinfo {author}
  {\bibfnamefont {C.}~\bibnamefont {Bittencourt}}, \ and\ \bibinfo {author}
  {\bibfnamefont {M.}~\bibnamefont {Prato}},\ }\href@noop {} {\bibfield
  {journal} {\bibinfo  {journal} {ACS Nano}\ }\textbf {\bibinfo {volume} {4}},\
  \bibinfo {pages} {6105} (\bibinfo {year} {2010})}\BibitemShut {NoStop}%
\bibitem [{\citenamefont {Sivaramakrishnan}\ \emph {et~al.}(2010)\citenamefont
  {Sivaramakrishnan}, \citenamefont {Wen}, \citenamefont {Scarpelli},
  \citenamefont {Pierce},\ and\ \citenamefont
  {Zuo}}]{Sivaramakrishnan:Equilibrium:2010}%
  \BibitemOpen
  \bibfield  {author} {\bibinfo {author} {\bibfnamefont {S.}~\bibnamefont
  {Sivaramakrishnan}}, \bibinfo {author} {\bibfnamefont {J.}~\bibnamefont
  {Wen}}, \bibinfo {author} {\bibfnamefont {M.~E.}\ \bibnamefont {Scarpelli}},
  \bibinfo {author} {\bibfnamefont {B.~J.}\ \bibnamefont {Pierce}}, \ and\
  \bibinfo {author} {\bibfnamefont {J.-M.}\ \bibnamefont {Zuo}},\ }\href
  {\doibase 10.1103/PhysRevB.82.195421} {\bibfield  {journal} {\bibinfo
  {journal} {Phys. Rev. B}\ }\textbf {\bibinfo {volume} {82}},\ \bibinfo
  {pages} {195421} (\bibinfo {year} {2010})}\BibitemShut {NoStop}%
\bibitem [{\citenamefont {Walter}\ \emph {et~al.}(2008)\citenamefont {Walter},
  \citenamefont {Akola}, \citenamefont {Lopez-Acevedo}, \citenamefont
  {Jadzinsky}, \citenamefont {Calero}, \citenamefont {Ackerson}, \citenamefont
  {Whetten}, \citenamefont {Groenbeck},\ and\ \citenamefont
  {H{\"a}kkinen}}]{ISI:000257645400007}%
  \BibitemOpen
  \bibfield  {author} {\bibinfo {author} {\bibfnamefont {M.}~\bibnamefont
  {Walter}}, \bibinfo {author} {\bibfnamefont {J.}~\bibnamefont {Akola}},
  \bibinfo {author} {\bibfnamefont {O.}~\bibnamefont {Lopez-Acevedo}}, \bibinfo
  {author} {\bibfnamefont {P.~D.}\ \bibnamefont {Jadzinsky}}, \bibinfo {author}
  {\bibfnamefont {G.}~\bibnamefont {Calero}}, \bibinfo {author} {\bibfnamefont
  {C.~J.}\ \bibnamefont {Ackerson}}, \bibinfo {author} {\bibfnamefont {R.~L.}\
  \bibnamefont {Whetten}}, \bibinfo {author} {\bibfnamefont {H.}~\bibnamefont
  {Groenbeck}}, \ and\ \bibinfo {author} {\bibfnamefont {H.}~\bibnamefont
  {H{\"a}kkinen}},\ }\href {\doibase {10.1073/pnas.0801001105}} {\bibfield
  {journal} {\bibinfo  {journal} {{P. Natl. Acad. Sci. USA}}\ }\textbf
  {\bibinfo {volume} {{105}}},\ \bibinfo {pages} {{9157}} (\bibinfo {year}
  {{2008}})}\BibitemShut {NoStop}%
\bibitem [{\citenamefont {Luedtke}\ and\ \citenamefont
  {Landman}(1998)}]{doi:10.1021/jp981745i}%
  \BibitemOpen
  \bibfield  {author} {\bibinfo {author} {\bibfnamefont {W.~D.}\ \bibnamefont
  {Luedtke}}\ and\ \bibinfo {author} {\bibfnamefont {U.}~\bibnamefont
  {Landman}},\ }\href {\doibase 10.1021/jp981745i} {\bibfield  {journal}
  {\bibinfo  {journal} {J. Phys. Chem. B}\ }\textbf {\bibinfo {volume} {102}},\
  \bibinfo {pages} {6566} (\bibinfo {year} {1998})}\BibitemShut {NoStop}%
\bibitem [{\citenamefont {Mpourmpakis}\ \emph {et~al.}(2010)\citenamefont
  {Mpourmpakis}, \citenamefont {Andriotis},\ and\ \citenamefont
  {Vlachos}}]{Mpourmpakis:Identification:2010}%
  \BibitemOpen
  \bibfield  {author} {\bibinfo {author} {\bibfnamefont {G.}~\bibnamefont
  {Mpourmpakis}}, \bibinfo {author} {\bibfnamefont {A.~N.}\ \bibnamefont
  {Andriotis}}, \ and\ \bibinfo {author} {\bibfnamefont {D.~G.}\ \bibnamefont
  {Vlachos}},\ }\href@noop {} {\bibfield  {journal} {\bibinfo  {journal} {Nano
  Lett.}\ }\textbf {\bibinfo {volume} {10}},\ \bibinfo {pages} {1041} (\bibinfo
  {year} {2010})}\BibitemShut {NoStop}%
\bibitem [{Note2()}]{Note2}%
  \BibitemOpen
  \bibinfo {note} {Sphericity equals $\pi ^{1/3}(6V)^{2/3}/A$ where $V$ is the
  volume and $A$ the area of the nanoparticle; characteristic values are 81\%
  for a cube, 85\% for an octahedron and 100\% for a sphere.}\BibitemShut
  {Stop}%
\bibitem [{\citenamefont {McKenna}\ and\ \citenamefont
  {Shluger}(2007)}]{mckenna07}%
  \BibitemOpen
  \bibfield  {author} {\bibinfo {author} {\bibfnamefont {K.~P.}\ \bibnamefont
  {McKenna}}\ and\ \bibinfo {author} {\bibfnamefont {A.~L.}\ \bibnamefont
  {Shluger}},\ }\href {\doibase 10.1021/jp710043s} {\bibfield  {journal}
  {\bibinfo  {journal} {J. Phys. Chem. C}\ }\textbf {\bibinfo {volume} {111}},\
  \bibinfo {pages} {18848} (\bibinfo {year} {2007})}\BibitemShut {NoStop}%
\bibitem [{\citenamefont {Lopez}\ \emph
  {et~al.}(2004{\natexlab{b}})\citenamefont {Lopez}, \citenamefont
  {N{\o}rskov}, \citenamefont {Janssens}, \citenamefont {Carlsson},
  \citenamefont {Puig-Molina}, \citenamefont {Clausen},\ and\ \citenamefont
  {Grunwaldt}}]{lopez04b}%
  \BibitemOpen
  \bibfield  {author} {\bibinfo {author} {\bibfnamefont {N.}~\bibnamefont
  {Lopez}}, \bibinfo {author} {\bibfnamefont {J.}~\bibnamefont {N{\o}rskov}},
  \bibinfo {author} {\bibfnamefont {T.}~\bibnamefont {Janssens}}, \bibinfo
  {author} {\bibfnamefont {A.}~\bibnamefont {Carlsson}}, \bibinfo {author}
  {\bibfnamefont {A.}~\bibnamefont {Puig-Molina}}, \bibinfo {author}
  {\bibfnamefont {B.}~\bibnamefont {Clausen}}, \ and\ \bibinfo {author}
  {\bibfnamefont {J.-D.}\ \bibnamefont {Grunwaldt}},\ }\href@noop {} {\bibfield
   {journal} {\bibinfo  {journal} {J. Catal.}\ }\textbf {\bibinfo {volume}
  {225}},\ \bibinfo {pages} {86} (\bibinfo {year}
  {2004}{\natexlab{b}})}\BibitemShut {NoStop}%
\end{thebibliography}%

\end{document}